# Dynamic glucose enhanced imaging using direct water saturation

Linda Knutsson[1,2,3] | Nirbhay N. Yadav[1,4] | Sajad Mohammed Ali[3] | David Olayinka Kamson[2,5] | Eleni Demetriou[1,4] | Anina Seidemo[6] | Lindsay Blair[2] | Doris D. Lin[4] | John Laterra[2,5,7,8] | Peter C. M. van Zijl[1,4,9]

[1]F.M. Kirby Research Center for Functional Brain Imaging, Kennedy Krieger Institute, Baltimore, Maryland, USA

[2]Department of Neurology, Johns Hopkins University School of Medicine, Baltimore, Maryland, USA

[3]Department of Medical Radiation Physics, Lund University, Lund, Sweden

[4]Russell H. Morgan Department of Radiology and Radiological Science, Johns Hopkins University School of Medicine, Baltimore, Maryland, USA

[5]Department of Oncology, Johns Hopkins University School of Medicine, Baltimore, Maryland, USA

[6]Diagnostic Radiology, Department of Clinical Sciences, Lund University, Lund, Sweden

[7]Hugo W. Moser Research Institute at Kennedy Krieger, Baltimore, Maryland, USA

[8]Department of Neuroscience, Johns Hopkins University School of Medicine, Baltimore, Maryland, USA

[9]Department of Biomedical Engineering, Johns Hopkins University School of Medicine, Baltimore, Maryland, USA

**Correspondence**
Linda Knutsson, F.M. Kirby Research Center, Kennedy Krieger Institute, 716 N. Broadway, Rm 131, Baltimore, MD, 21205, USA.
Email: lknutss1@jhu.edu

**Funding information**
National Institute of Biomedical Imaging and Bioengineering, Grant/Award Numbers: RO1 EB034978, S10OD021648; Vetenskapsrådet, Grant/Award Number: 2019-03637; Cancerfonden, Grant/Award Number: 21 1652 Pj

**Abstract**

**Purpose:** Dynamic glucose enhanced (DGE) MRI studies employ CEST or spin lock (CESL) to study glucose uptake. Currently, these methods are hampered by low effect size and sensitivity to motion. To overcome this, we propose to utilize exchange-based linewidth $(LW)$ broadening of the direct water saturation (DS) curve of the water saturation spectrum (Z-spectrum) during and after glucose infusion (DS-DGE MRI).

**Methods:** To estimate the glucose-infusion-induced $LW$ changes $(\Delta LW)$, Bloch-McConnell simulations were performed for normoglycemia and hyperglycemia in blood, gray matter (GM), white matter (WM), CSF, and malignant tumor tissue. Whole-brain DS-DGE imaging was implemented at 3 T using dynamic Z-spectral acquisitions (1.2 s per offset frequency, 38 s per spectrum) and assessed on four brain tumor patients using infusion of 35 g of D-glucose. To assess $\Delta LW$, a deep learning-based Lorentzian fitting approach was used on voxel-based DS spectra acquired before, during, and post-infusion. Area-under-the-curve $(AUC)$ images, obtained from the dynamic $\Delta LW$ time curves, were compared qualitatively to perfusion-weighted imaging parametric maps.

**Results:** In simulations, $\Delta LW$ was 1.3%, 0.30%, 0.29/0.34%, 7.5%, and 13% in arterial blood, venous blood, GM/WM, malignant tumor tissue, and CSF, respectively. In vivo, $\Delta LW$ was approximately 1% in GM/WM, 5% to 20% for different tumor types, and 40% in CSF. The resulting DS-DGE $AUC$ maps clearly outlined lesion areas.

**Conclusions:** DS-DGE MRI is highly promising for assessing D-glucose uptake. Initial results in brain tumor patients show high-quality $AUC$ maps of glucose-induced line broadening and DGE-based lesion enhancement similar and/or complementary to perfusion-weighted imaging.

## 1 | INTRODUCTION

Gadolinium (Gd)-based contrast agents (GBCA) play a major role in MRI for both research and clinical routine. However, their use in certain patient groups is limited due to side effects such as nephrogenic systemic fibrosis.[1,2] In vivo deposition[3–5] has also led to the United States Food and Drug Administration (FDA) issuing a box package warning on GBCAs, which remain under continued review.[6] In addition, many malignant tumors show little to no Gd-enhancement.[7,8] Consequently, current clinical practice is judicious use of GBCA, particularly in young and vulnerable populations.[9] Therefore, there is a need to develop new contrast agents. The availability of chemical exchange saturation transfer (CEST)[10–17] and chemical exchange sensitive spin lock (CESL)[18–29] approaches has opened the MRI field to non-metallic contrast agents. When D-glucose (D-Glc) is used as an agent, these approaches have been dubbed glucoCEST and glucoCESL, respectively.

Dynamic glucose enhanced (DGE) MRI applies CEST or CESL of sugars dynamically, providing information on contrast agent uptake in tissues in a manner similar to dynamic contrast enhanced (DCE) and dynamic susceptibility contrast (DSC) MRI.[13,14,23–25,27,28,30–35] Unfortunately, DGE MRI signal changes at clinical field strengths (3 T) have been on the order of 1%,[17,32] resulting in sensitivity to motion artifacts, especially when based on a single signal intensity from one saturation offset frequency per dynamic.[36,37] Here, we propose to reduce these problems by utilizing the transverse relaxation effect originating from the chemical shift difference between the hydroxyl and water proton pools. The exchange between these pools leads to spins experiencing different precession frequencies,[38,39] resulting in a collective phase dispersion and a linewidth $(LW)$ broadening of the direct saturation (DS) line shape in water saturation spectra (Z-spectra). We exploit this exchange-based relaxation enhancement to assess changes in D-Glc concentration by acquiring DS spectra using RF saturation of short duration and low $B_1$. This approach is commonly used in the water saturation shift referencing (WASSR) method[40] for measuring $B_0$ shifts, since it minimizes contributions from semi-solid magnetization transfer contrast (MTC), CEST, and relayed nuclear Overhauser effects (rNOEs).[41] This allows fitting of the full DS spectrum to a Lorentzian,[42] a method recently optimized using deep learning (DL).[43]

This study, therefore, aimed to develop a WASSR-analogous DGE method to utilize glucose-induced increases in the DS linewidth, dubbed DS-DGE MRI. Its feasibility in different types of tissues at 3 T was investigated through simulations and for D-Glc infusions in brain tumor patients. The in vivo DS-DGE effect size was compared qualitatively with DCE- and DSC-MRI parametric maps to investigate whether similar and/or complementary information was obtained.

## 2 | METHODS

### 2.1 | Simulations

Bloch-McConnell simulations of Z-spectra at 3 T before and after D-Glc infusion were performed using Pulseq-CEST.[44] Five tissues were simulated: blood, gray matter (GM), white matter (WM), malignant tumor with blood–brain barrier (BBB) disruption (TUMOR), and CSF. Importantly, hydroxyl exchange properties and water transverse relaxation times may differ between tissue compartments, leading to different signal contributions. Therefore, following a recently established model,[45] three tissue compartments were simulated within WM, GM and TUMOR, namely blood (b), extravascular extracellular space (EES), and cell (c). Within blood, we assumed arteriolar (a) and venular (v) compartments. A previous DGE MRI study reported a venous plasma blood glucose average increase of 9.8 mM (n = 11) for a D-Glc dose of 25 g.[30] Since our experiments used a 40% higher dose (35 g), we assumed a 13.7 mM increase in venous blood D-Glc concentration from normoglycemic baseline to hyperglycemia. The resulting compartmental D-Glc concentrations after transport[45] (Table 1) were used to calculate the Z-spectral intensities, $\frac{S(\Delta\omega)}{S_0}$, using the hydroxyl proton pools at 0.66, 1.28, 2.08, and 2.88 ppm.[48] All compartments, except for TUMOR EES, were assumed to have a pH of 7.2, with hydroxyl proton exchange rates of 2900, 6500, 5200, and 14 300 Hz, respectively,[48] at 37°C. For TUMOR EES, a pH of 6.8 was assumed, leading to exchange rates of 1500, 3100, 2500, and 6000 Hz, respectively.[48] Compartmental Z-spectra were simulated using 41 frequency offsets: ±[10, 5.0, 4.0, 3.0, 2.5, 2.0, 1.75, 1.5, 1.25, 1.0, 0.80, 0.65, 0.55, 0.48, 0.40, 0.33, 0.26, 0.18, 0.11, 0.036, 0.0] ppm. Saturation was applied using 10 consecutive 50-ms sinc-Gaussian pulses, resulting in a total saturation time ($t_{sat}$) of 0.5 s. The simulations were performed using a $B_{1peak}$ of 0.5 μT.

Z-spectra in GM, WM, and TUMOR were calculated by adding the normalized signals of blood, EES, and cell[45]:

$$\frac{S_{tis}(\Delta\omega)}{S_0} = \left[ f_{b,tis}\rho_b^{V/V} \left\{ f_a \frac{S_a(\Delta\omega)}{S_{0a}} + f_v \frac{S_v(\Delta\omega)}{S_{0v}} \right\} + f_{e,tis}\rho_e^{V/V} \frac{S_{e,tis}(\Delta\omega)}{S_{0e}} + f_{c,tis}\rho_{c,tis}^{V/V} \frac{S_{c,tis}(\Delta\omega)}{S_{0c}} \right] / \rho_{tis}^{V/V}, \quad (1)$$

with $f_{i,tis}$ being the volume fraction for compartment $i$ in mL compartment/mL tissue (Table 1). Capillary blood was assumed to have fast deoxygenation, resulting in the blood

**TABLE 1** Relaxation times,* fractional volumes, D-Glc concentrations, and water contents for tissues (b, GM, WM, TUMOR, and CSF) and their compartments (arterial, venous, EES, cell).

| Parameter | Tissue | | | | | |
|---|---|---|---|---|---|---|
| | Arterial blood** | Venous blood** | GM | WM | TUMOR | CSF |
| $T_{1,b}$ (s) | 1.91 | 1.73 | | | | |
| $T^e_{1,tis}$ (s) | | | 3.48 | 3.48 | 3.48 | 3.48 |
| $T^c_{1,tis}$ (s) | | | 1.08 | 0.65 | 1.02 | |
| $T_{2,b}$ (s) | 0.152 | 0.052 | | | | |
| $T^e_{2,tis}$ (s) | | | 2.78 | 2.78 | 2.78 | 2.78 |
| $T^c_{2,tis}$ (s) | | | 0.071 | 0.055 | 0.071 | |
| $f_{comp,b}$ (mL comp/mL b) | 0.30 | 0.70 | | | | |
| $f_{b,tis}$ (mL b/mL tis) | | | 0.038 | 0.018 | 0.050 | 0.00 |
| $f_{e,tis}$ (mL EES/mL tis) | | | 0.22 | 0.22 | 0.50 | 1.00 |
| $f_{c,tis}$ (mL cell/mL tis) | | | 0.74 | 0.76 | 0.45 | 0.00 |
| $C_b$ ngl (mM) | 6.15 | 5.47 | | | | |
| $C_{e,tis}$ ngl (mM) | | | 2.24 | 2.42 | 5.45 | 3.69 |
| $C_{c,tis}$ ngl (mM) | | | 0.167 | 0.381 | 0.811 | |
| $C_b$ hgl (mM) | 19.8 | 19.1 | | | | |
| $C_{e,tis}$ hgl (mM) | | | 6.10 | 5.71 | 17.6 | 11.9 |
| $C_{c,tis}$ hgl (mM) | | | 1.16 | 1.24 | 4.41 | |
| $\rho^{V/V}_b$ (mL water/mL b) | | | 0.856 | 0.856 | 0.856 | |
| $\rho^{V/V}_e$ (mL water/mL comp) | | | 0.938 | 0.938 | 0.938 | |
| $\rho^{V/V}_c$ (mL water/mL comp) | | | 0.809 | 0.678 | 0.674 | |
| $\rho^{V/V}_{tis}$ (mL water/mL tis) | | | 0.839 | 0.738 | 0.815 | |

*Note*: Values for $f$, $C$, and $\rho^{V/V}$ are from Seidemo et al.[45] and references therein.

Abbreviations: *a*, arterial; *b*, blood; *c*, cell; *C*, D-Glc concentration; *comp*, compartment; *D-Glc*, D-glucose; *e*, EES; *EES*, extravascular extracellular space; *f*, volume fraction ($f_a + f_v = 1, f_{b,tis} + f_{e,tis} + f_{c,tis} = 1$); *GM*, gray matter; *hgl*, hyperglycemia; *ngl*, normoglycemia; $\rho^{V/V}$, water volume density; *tis*, tissue; *TUMOR*, malignant tumor with blood–brain barrier disruption; *v*, venous; *WM*, white matter.

*OH was set to have a $T_1$ of 1.20 s and $T_2$ of 0.100 s.

**For arterial and venous blood, a well-mixed compartment (fast exchange for water between plasma and erythrocytes) was assumed for the 500 ms saturation period. Relaxation times were calculated based on references,[46,47] using: https://www.kennedykrieger.org/physiologic-metabolic-anatomic-biomarkers/resources/software-and-databases/blood-t2-t1-hct-and-oxygenation-calculator and the following settings: 3 T; $t_{cp}$ 50 ms; $t_p$ 50 ms; Hematocrit (Hct) 0.4; Oxygenations (Y) 0.98 and 0.60 for arterial and venous blood, respectively.

compartment (b) consisting of only arteriolar (a) and venular (v) subcompartments[49] with volume fractions $f_a = 0.3$ and $f_v = 0.7$ in mL/mL blood, respectively. Since MR contrast is determined by water-based compartmental concentrations and tissue volume fractions, corrections are included for $\rho^{V/V}$, the water content per mL compartment or tissue (Table 1). After adding the Z-spectra from each compartment, 2% Rician noise was applied. The resulting tissue Z-spectra were re-sampled at frequencies used in the experiments (28 frequencies from −5 to 5 ppm), followed by Lorentzian fitting using deep-learning.[43] Thereafter, the linewidth difference ($\Delta LW$) between normoglycemia (baseline) and hyperglycemia was calculated for each tissue using Eq. 2 below.

## 2.2 | Experiments

### 2.2.1 | Patients

Four brain tumor patients were studied (one with an isocitrate dehydrogenase (IDH)-wildtype glioblastoma, two with a grade 2 IDH-mutated astrocytoma, and one with anaplastic lymphoma kinase (ALK)-mutated non–small-cell lung cancer metastases). The project was approved by the local Institutional Review Board, and each participant provided written informed consent. Participants were asked to fast 6 h before the study, but clear liquids were permitted. Before the start of the MRI examination, blood was drawn to verify normal baseline glucose levels (3.9–7.0 mM). Supporting Information (Table S1) lists the study's exclusion criteria.

### 2.2.2 | MRI acquisition protocol

Patients were examined on a 3 T Philips Elition RX system (Philips Healthcare). Pre- and post-contrast enhanced $T_1$-MPRAGE, fluid-attenuated inversion recovery (FLAIR), DS-DGE, DCE and DSC images were acquired. $T_1$-MPRAGE: TR/TE/flip angle (FA) = 7.5 ms/3.5 ms/8°, FOV = 212 × 212 mm$^2$, resolution = 1.1 × 1.1 × 2.2 mm$^3$, inversion time = 755 ms, acquisition time = 1 min 46 s, and FLAIR: TR/TE = 11 000/120 ms, FOV = 212 × 212 mm$^2$, resolution = 1.1 × 1.1 × 2.2 mm$^3$, inversion time = 2800 ms, acquisition time = 3 min 51 s.

DS-DGE images were acquired using 10 consecutive 50-ms sinc-Gaussian pulses ($B_{1peak}$ = 0.5 μT, $t_{sat}$ = 0.5 s), followed by a whole-brain simultaneous multi-slice EPI readout (multi-band factor 3). A total of 27 slices with FOV of 208 × 208 mm$^2$ and a resolution of 2.2 × 2.2 × 4.4 mm$^3$ were acquired using TR/TE/FA = 1200 ms/17 ms/52°. Thirty-two frequencies were acquired in 38.2 s at offsets: ±[10 (2×), 5.0, 2.5, 2.0, 1.5, 1.2, 1.0, 0.80, 0.70, 0.60, 0.50, 0.40, 0.30, 0.20, 0.10] ppm. In total 40 dynamics (Z-spectra) were acquired. Approximately 5 min into the scan (after the eighth dynamic), D-Glc was administrated intravenously with a power-injector at a rate of 6.25 g/min using hospital-grade D50 glucose (D50, Hospira; 35 g of D-Glc in 70 mL of water sterile solution prepared by the Johns Hopkins pharmacy), followed by a saline rinse. Total experiment duration was 25.5 min.

A $T_1$-weighted gradient-echo sequence was used for DCE MRI: TR/TE/FA = 5.1 ms/2.5 ms/26°, FOV = 212 × 212 mm$^2$, resolution = 2.2 × 2.2 × 4.4 mm$^3$. Gadoteridol (ProHance, Bracco Diagnostics, 0.1 mmol/kg) was given at a rate of 5 mL/s via a power injector followed by a saline rinse. The injection delay was 30 s (15 pre-contrast baseline images). Each 15-slice dynamic scan was 2.0 s and a total of 150 dynamics over 5 min was acquired. After the DCE, post-contrast MPRAGE images were obtained.

Approximately 7 min after the first GBCA injection, a second GBCA dose was injected at a 5 mL/s rate via a power injector, followed by a saline rinse. DSC was performed using single-shot EPI with TR/TE/FA = 1344 ms/29 ms/90°, FOV = 212 × 212 mm$^2$, and resolution = 2.2 × 2.2 × 4.4 mm$^3$. A 15-s injection delay (11 pre-contrast baseline images) was set. A total of 100 dynamics were acquired over 2.2 min for 25 slices.

## 2.3 | Post-processing

### 2.3.1 | DS-DGE MRI

The dynamic DS spectrum signal intensities from each voxel were normalized using the average of the second of two acquisitions at ±10 ppm and then fitted to a Lorentzian line shape using the DL-based single Lorentzian fitting neural network.[43] The *LWs*, defined as *FWHM* in the DL fitting, were used to generate *LW* maps (in Hz) for each dynamic. The dynamic *LW* maps were rigid motion corrected using Elastix[50] and visually inspected for remaining motion artifacts.

A baseline was calculated by averaging the *LW* maps obtained before infusion. To remove outliers, baseline *LWs* greater or less than the average ± 2SD were discarded. $LW_{base}$ was then calculated by averaging the remaining baseline *LWs*. Dynamic Δ*LW* images were obtained by subtracting $LW_{base}$ from each *LW* dynamic image, *LW(t)*, followed by normalization with $LW_{base}$:

$$\Delta LW(t)\,(\%) = \frac{LW(t) - LW_{base}}{LW_{base}} \times 100\%. \quad (2)$$

Normalized area-under-the-curve *(AUC)* was calculated by subtracting $LW_{base}$ from the average of dynamic *LWs* obtained from infusion start and through the dynamic scan ($LW_{average}$), followed by normalization with $LW_{base}$:

$$AUC\,(\%) = \frac{LW_{average} - LW_{base}}{LW_{base}} \times 100\%. \quad (3)$$

Normalized *AUC* over the infusion block alone was also calculated. Both fitting and calculations were performed in Python.

### 2.3.2 | DCE and DSC MRI

DCE and DSC MRI were processed using OLEA Sphere (Olea Medical Solutions). For both sequences, motion correction was applied before the tracer kinetic modeling. The extended Toft model was used for DCE-MRI[51] to retrieve interstitial volume ($V_e$) and the volume transfer constant ($K^{trans}$), a measure combining permeability and perfusion. For DSC-MRI, standard tracer kinetic modeling including leakage correction was applied to obtain leakage corrected cerebral blood volume *(corr. CBV)*, uncorrected cerebral blood volume *(uncorr. CBV)* and leakage *(K2)*.[52,53]

## 3 | RESULTS

Figure 1 shows the simulated Z-spectra for the different tissues with and without glucose infusion. Simulated baseline $LWs$ were 87 Hz for blood, 57 Hz for arterial blood, 96 Hz for venous blood, 65 Hz for GM, 60 Hz for WM, 42 Hz for TUMOR, and 16 Hz for CSF. Simulated $\Delta LWs$ were 0.56% for blood, 1.3% for arterial blood, 0.30% for venous blood, 0.29% for GM, 0.34% for WM, 7.5% for TUMOR, and 13% for CSF.

Figure 2 shows a patient with a recurrent IDH-wildtype glioblastoma with thin peripheral contrast-enhancement (Gd-$T_1$w) around the resection cavity and surrounding expansile infiltrative FLAIR hyperintense tumor. Dynamic $\Delta LW$ images obtained on infusing D-Glc are shown (averaged over two dynamics). Note the $LW$ increases in vascular, CSF, and tumor tissue. For DCE, $K^{trans}$ shows an increase in the same regions, while $V_e$ shows only a slight increase. Both $K^{trans}$ and $V_e$ are hypointense inside the cavity. The DS-DGE *AUC* map also displays an increase in the surrounding tumor and a hypointense core. For DSC, $K2$ shows enhancement comparable to DS-DGE. During the D-Glc infusion, the *LW* increased to approximately 15% for the contrast enhanced peri-cavity infiltrative tumor region and 40% for ventricular CSF. Figure 3 shows experimental Z-spectra before and after D-Glc infusion together with DL Lorentzian fits for region-of-interests (ROIs) placed in the DS-DGE *AUC*, with assistance from Gd-$T_1$w and FLAIR images, in the anterior cerebral artery, GM, WM, tumor tissue, and CSF (ventricle). Notice that the *LWs* and their broadening are of a similar order of magnitude to those simulated in Figure 1. However, the in vivo changes in arterial blood and CSF were generally larger than those in the simulations.

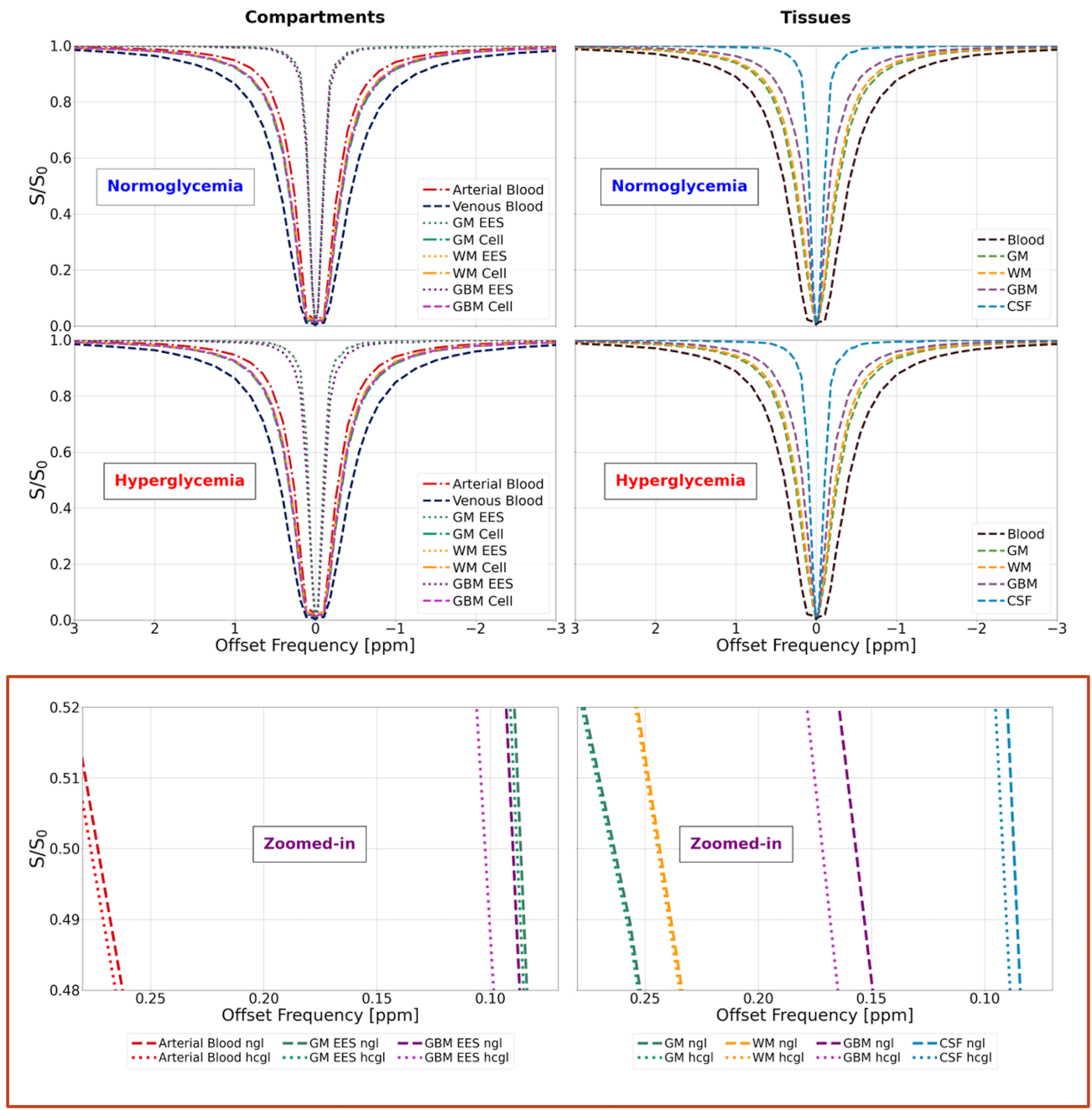

**FIGURE 1** Normoglycemic ($C_p^a = 6.15$ mM) and hyperglycemic ($C_p^a = 19.8$ mM) simulated Z-spectra for tissue compartments (left) and total tissue (right). The lower row shows a zoomed-in view comparing the Z-spectral intensities around half-maximum for normoglycemia and hyperglycemia. Only Z-spectra with a sufficiently large change are visualized for tissue compartments (lower left). Saturation parameters: $B_{1peak} = 0.5\ \mu T$, 10 consecutive 50-ms sinc-Gauss pulses for $t_{sat} = 0.5$ s (TR = 1.2 s).

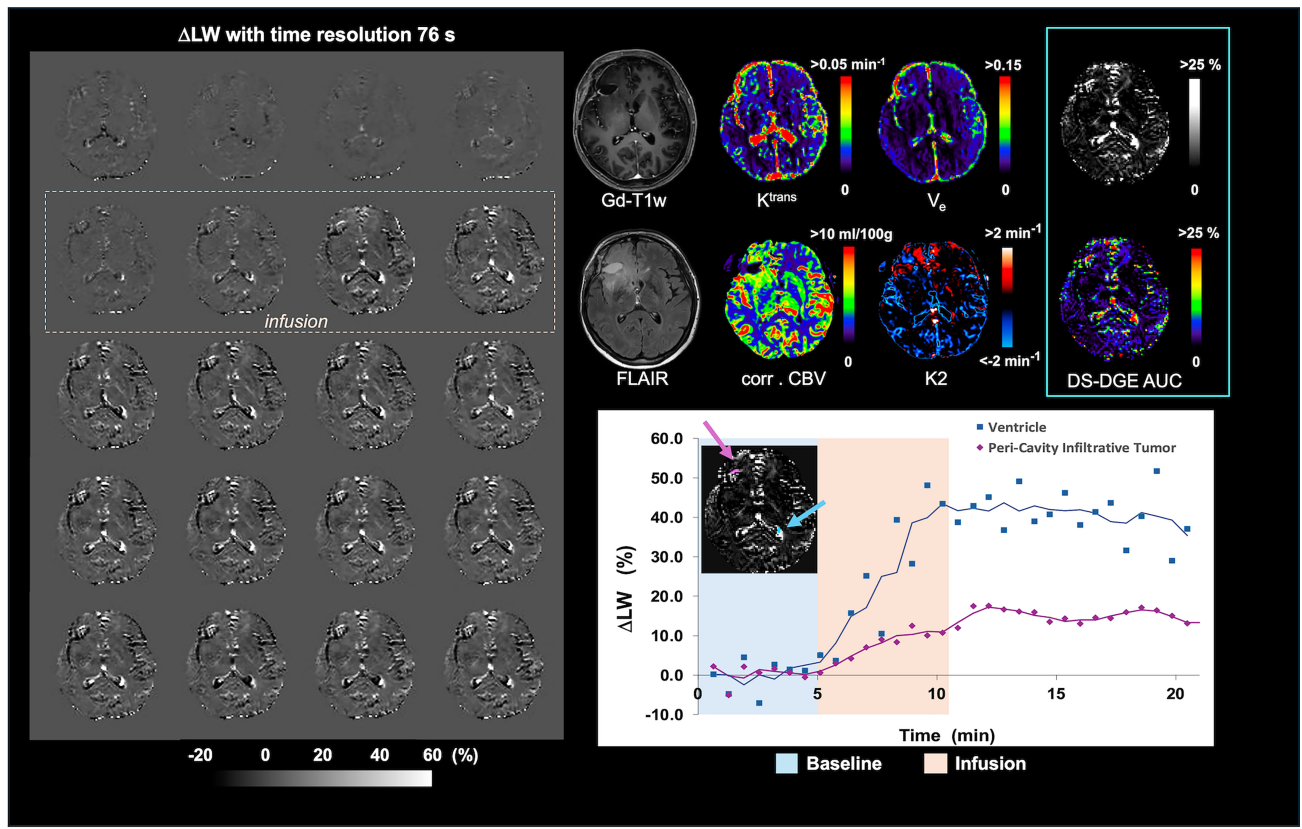

**FIGURE 2** Patient with recurrent IDH-wildtype glioblastoma showing thin Gd-enhancement around the resection cavity. (Left) $\Delta LW$ maps during the scan (averaged over a period of 76 s corresponding to two $\Delta LW$ images). (Right top) Anatomical images (Gd-$T_1$w, FLAIR), together with parametric maps from DCE MRI ($K^{trans}$, $V_e$), DSC MRI (*corr. CBV*, $K2$) and DS-DGE MRI (*AUC* grayscale and color-coded). (Right bottom) Graph of linewidth change versus time obtained from regions of interest (ROIs) placed in the DS-DGE peri-cavity infiltrative tumor region, located anterior to the cavity, and ventricle (purple diamonds and blue squares, respectively). The ROIs are overlaid on the DS-DGE MRI *AUC* map located in the graph as purple and blue areas, respectively. To visualize the trend in glucose uptake, $\Delta LW(t)$ curves were temporally smoothed with a 3-point moving average (purple and blue lines). $\Delta LW$, glucose-infusion-induced *LW* change; *AUC*, area-under-curve; *corr. CBV*, corrected cerebral blood volume; *DCE*, dynamic contrast enhanced; *DSC*, dynamic susceptibility contrast; *DS-DGE*, direct water saturation-dynamic glucose enhanced; *FLAIR*, fluid-attenuated inversion recovery; *Gd*, gadolinium; *K2*, leakage; $K^{trans}$, volume transfer constant; *LW*, linewidth, $T_1$w, $T_1$ weighted; $V_e$, interstitial volume.

Figure 4 shows a patient with a grade 2 IDH-mutated astrocytoma. Interestingly, the enhanced tumor rim in the DS-DGE *AUC* image corresponds approximately to the spatial difference between the hypointense area in Gd-$T_1$w and the hyperintense area in FLAIR. The enhanced areas during infusion only and over the experimental duration are of comparable size. Corrected *CBV* and $K2$ also show an increase in the corresponding area. However, $K^{trans}$ and $V_e$ appear normal.

Figure 5 shows four slices from a patient with a grade 2 IDH-mutated astrocytoma. $K^{trans}$ shows an increase in the tumor boundary, while $V_e$ shows only a slight increase. Similar to the patient in Figure 1, the *CBV* lesion is strongly reduced in intensity and area after leakage correction, as reflected in the $K2$ enhancement. Note that DS-DGE hyperintense and hypointense tumor regions correspond approximately to the hyperintense rims and hypointense cores in the FLAIR images,

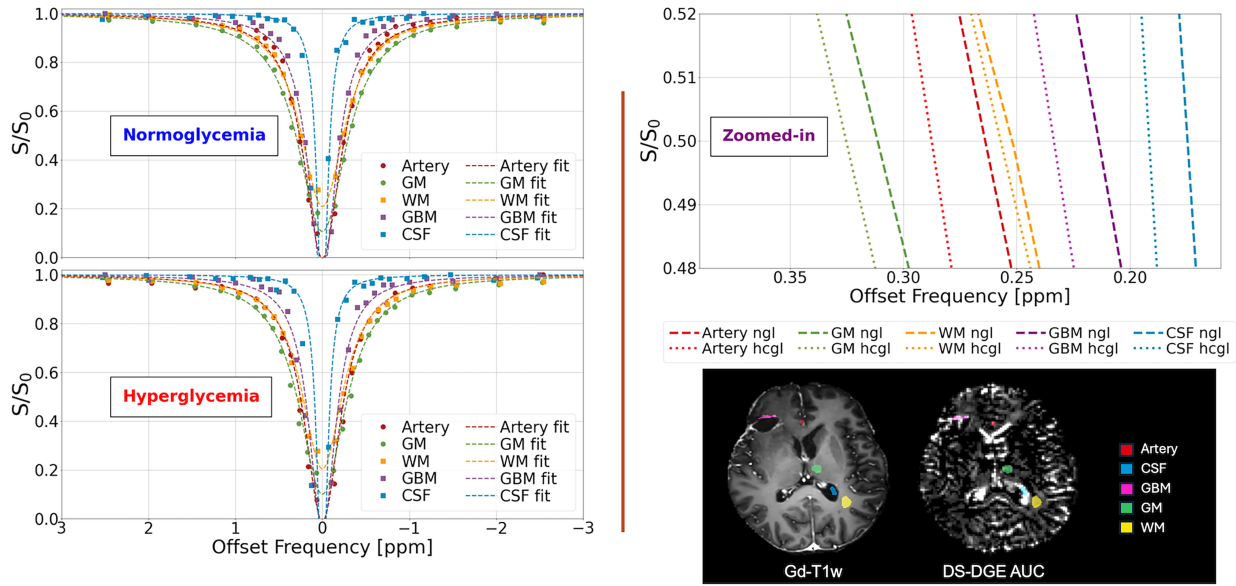

**FIGURE 3** (Left) Normoglycemic and hyperglycemic experimental Z-spectra from the glioblastoma patient in Figure 2 and the corresponding DL Lorentzian fits. (Right) A zoomed-in view demonstrating the linewidth difference between normoglycemic and hyperglycemic experimental Z-spectra. ROI locations are displayed in the Gd-$T_1$w image and DS-DGE *AUC* map. *AUC*, area-under-curve; *DS-DGE*, direct water saturation-dynamic glucose enhanced; *Gd*, gadolinium; *hcgl*, hyperglycemic; *ngl*, normoglycemic; $T_1$w, $T_1$ weighted.

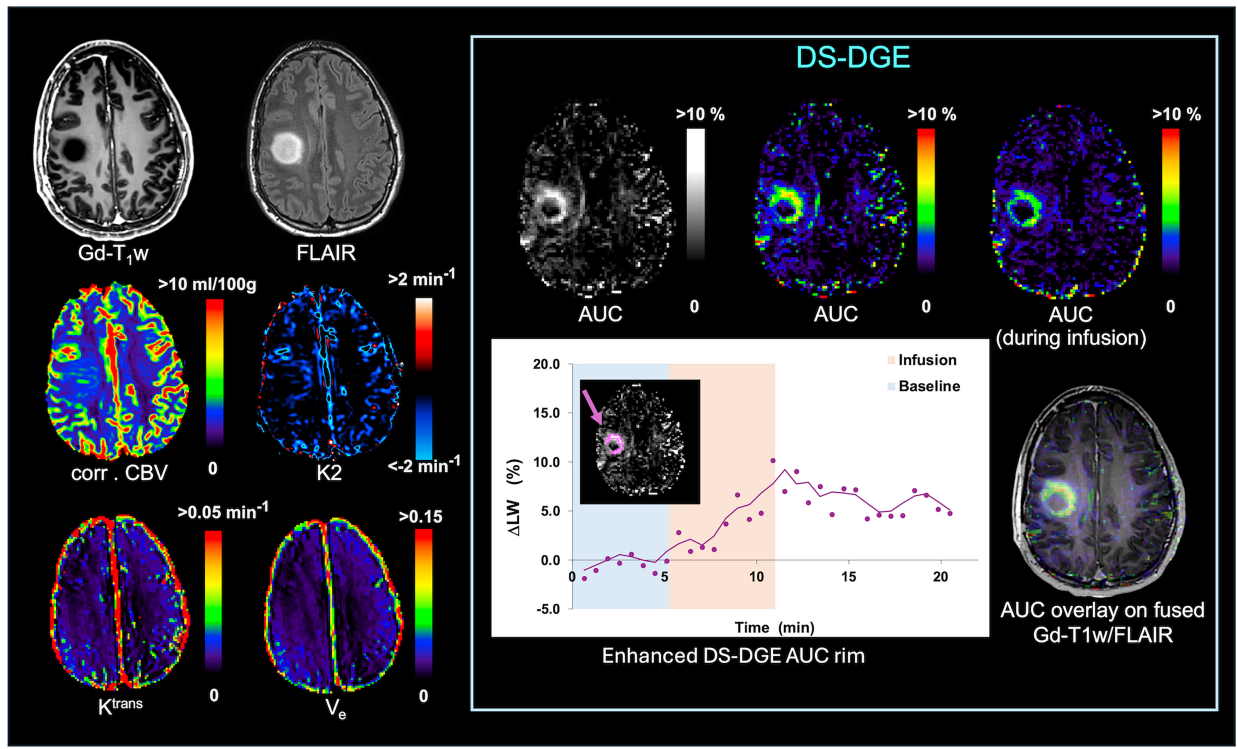

**FIGURE 4** Patient with grade 2 IDH-mutated astrocytoma. Anatomical images (Gd-$T_1$w, FLAIR) together with *corr. CBV, K2, $K^{trans}$, $V_e$*, and DS-DGE MRI maps (*AUC* in both grayscale and color-coded). Color-coded *AUC* calculated from the infusion block only is also shown. A DS-DGE *AUC* map overlayed on fused Gd-$T_1$w/FLAIR is shown for reference. Graph of linewidth change versus time obtained from region of interest (ROI) placed in the DS-DGE contrast-enhanced area (purple overlayed on DS-DGE *AUC* map). To visualize the trend in glucose uptake, Δ*LW(t)* curves (purple dots) were temporally smoothed with a 3-point moving average (purple line). Δ*LW*, glucose-infusion-induced *LW* change; *AUC*, area-under-curve; *corr. CBV*, corrected cerebral blood volume; *DS-DGE*, direct water saturation-dynamic glucose enhanced; *FLAIR*, fluid-attenuated inversion recovery; *Gd*, gadolinium; *K2*, leakage; $K^{trans}$, volume transfer constant; *LW*, linewidth; $T_1$w, $T_1$ weighted; $V_e$, interstitial volume.

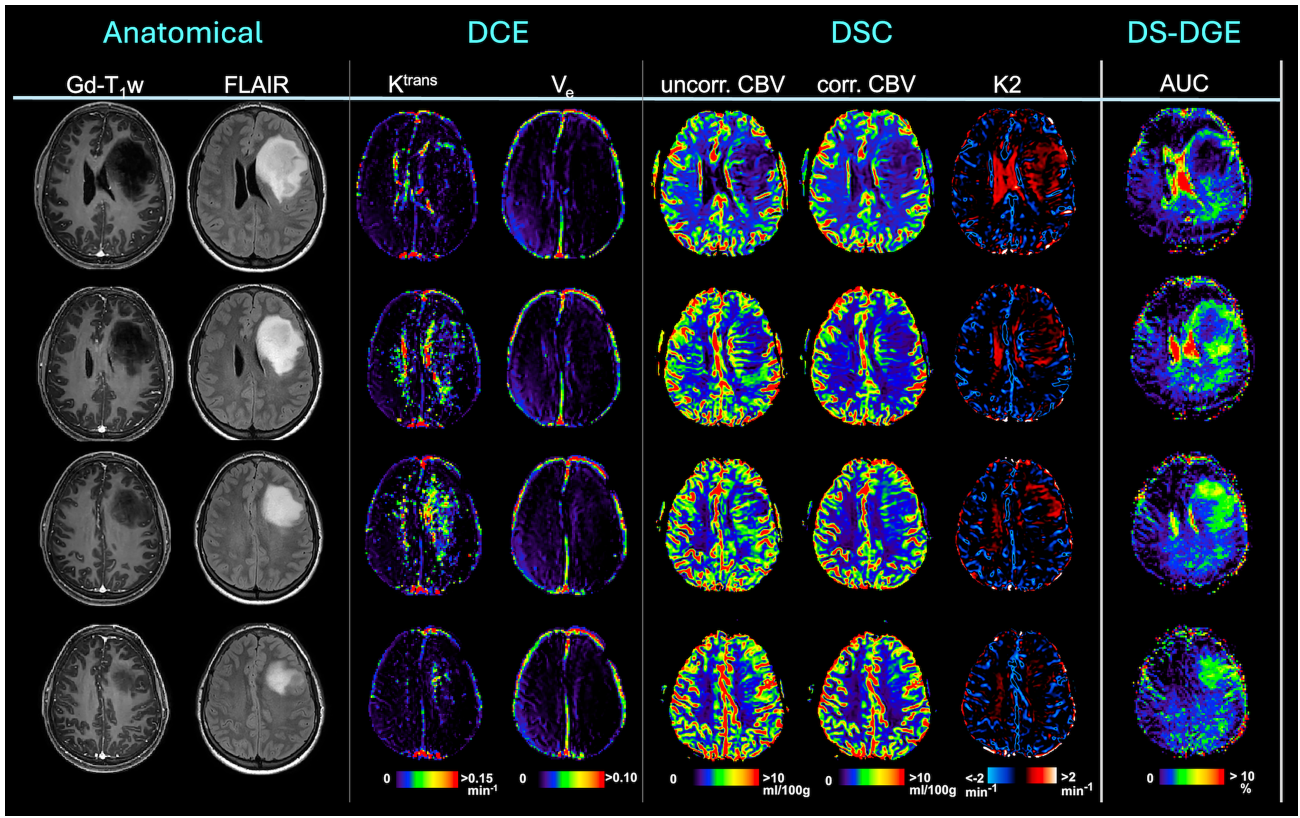

**FIGURE 5** Patient with a grade 2 IDH-mutated astrocytoma. Anatomical images (Gd-$T_1$w, FLAIR) together with parametric maps from DCE MRI ($K^{trans}$, $V_e$), DSC MRI (*uncorr. CBV, corr. CBV, K2*) and DS-DGE MRI (color-coded *AUC*) from four slices. *AUC*, area-under-curve; *corr. CBV*, corrected cerebral blood volume; *DCE*, dynamic contrast enhanced; *DSC*, dynamic susceptibility contrast; *DS-DGE*, direct water saturation-dynamic glucose enhanced; *FLAIR*, fluid-attenuated inversion recovery; *Gd*, gadolinium; *K2*, leakage; $K^{trans}$, volume transfer constant; $T_1$w, $T_1$ weighted; *uncorr. CBV*, uncorrected cerebral blood volume; $V_e$, interstitial volume.

respectively. The *K2* images show a similar trend, but over a smaller area.

Figure 6 shows results for a patient with ALK-mutated non–small-cell lung cancer brain metastases. $K^{trans}$ and $V_e$ are increased in the Gd-contrast-enhanced lesion area. The DS-DGE *AUC* map also displays an increase in part of the lesion area, while the WM shows negligible *LW* change. Note that leakage correction strongly reduced the elevated uncorrected *CBV* values in the lesion area. The enhancement observed on the contralateral side in the DS-DGE map is due to ventricular CSF. The Δ*LW* time curve shows a continuous increase up to approximately 20% in the DS-DGE contrast-enhanced lesion, resulting in a relatively smaller lesion area enhancement in the DS-DGE *AUC* map calculated from the infusion block only.

## 4 | DISCUSSION

We developed and implemented DS-DGE MRI to dynamically assess D-Glc uptake in brain tumors. This approach samples Z-spectra dynamically, which has the advantages of (1) multiple signal points per dynamic (leading to higher SNR and reduced motion sensitivity due to the Lorentzian fitting); (2) being independent of $B_0$ changes in the voxel between dynamics, for example, such as those due to motion, as the full DS spectrum is fitted; and (3) a water resonance line shape that has minimal contributions from CEST, rNOE, and MTC effects and can be approximated by a Lorentzian curve.[40,42,54–56] When applying RF saturation with low $B_1$, the FWHM of the DS Z-spectral line can be calculated as:

$$LW = FWHM = \left(\frac{1}{\pi}\right)\sqrt{\frac{R_1 R_2^2 + \omega_1^2 R_2}{R_1}} \approx \left(\frac{1}{\pi}\right)\sqrt{\frac{\omega_1^2 R_2}{R_1}}, \quad (4)$$

in which $\omega_1 = \gamma B_1$ (in units of rad/s) and $R_{1,2} = 1/T_{1,2}$. The presence of exchangeable protons at an offset from the water resonance will increase $R_2$[38,39] and broaden this line shape, with relative effects expected to be largest for tissue compartments with a long water $T_2$. This was borne out in the simulations, where EES and CSF exhibited narrow lines and large *LW* changes when the D-Glc concentration was increased (Figure 1). When compartments were added proportionally (Eq. [1]) to generate the Z-spectra for GM, WM, and TUMOR, Δ*LW* was highest in TUMOR, which is attributed to (1) the large concentration of D-Glc in EES after BBB breakdown; (2) increased blood volume and EES volume; and (3) the lower pH in tumor EES, which reduces the exchange rate and increases the exchange-based transverse relaxivity. For blood, despite having the highest D-Glc concentration, simulations showed a smaller *LW* change, which we attribute to the shorter $T_2$ originating from the high protein contents in both plasma (albumin) and erythrocyte (hemoglobin). Notice that an equal water-based concentration of D-Glc in plasma and erythrocytes[57–59] was used, removing any effect of microvascular hematocrit on blood D-Glc concentrations. The small effects in WM and GM are attributed to the low D-Glc concentration in EES and even lower in the cells due to facilitated transport over the BBB and the cell membrane, respectively, and metabolism in the cells. The short $T_2$ of the cell compartment, which occupies a large volume fraction of the tissue, further reduces the DS-DGE effect size.

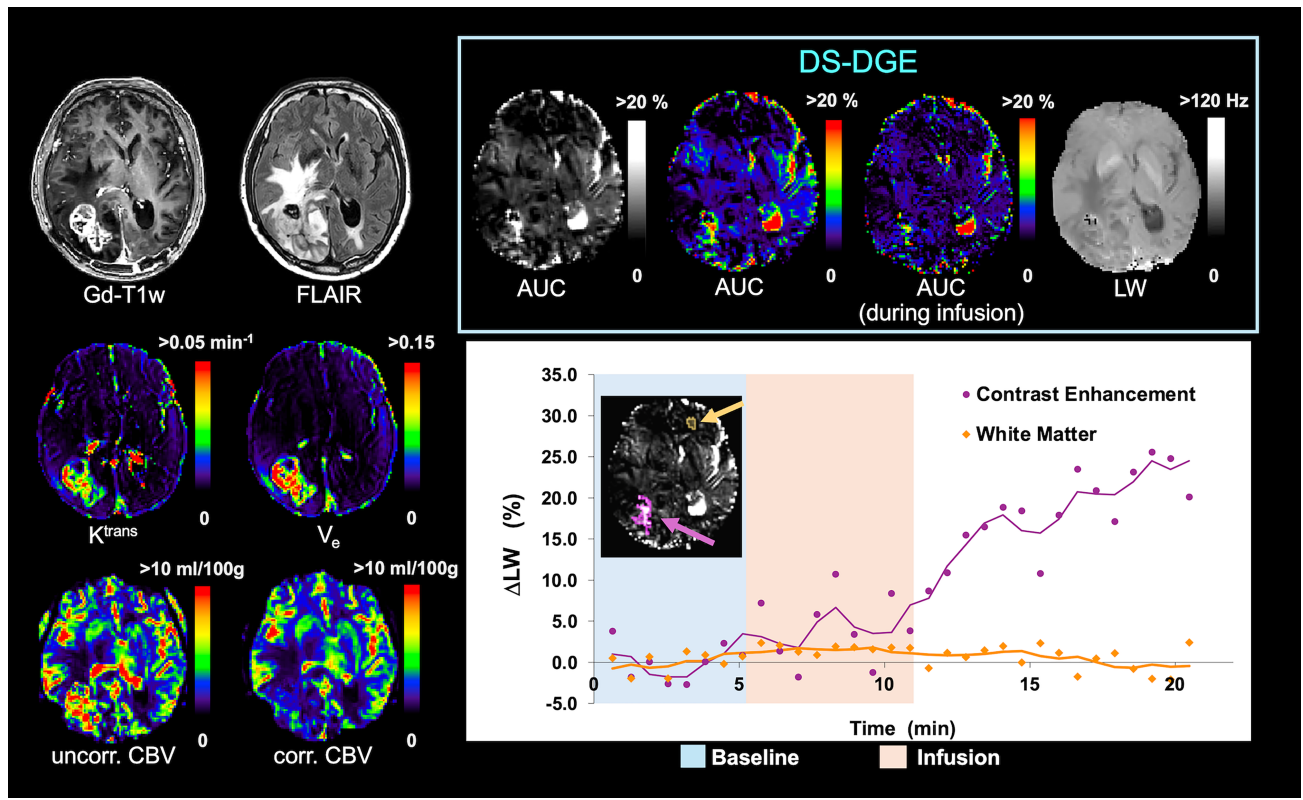

**FIGURE 6** Patient with brain metastasis from anaplastic lymphoma kinase-mutated non–small-cell lung cancer. Anatomical images (Gd-T$_1$w, FLAIR) together with $K^{trans}$, $V_e$, *uncorr. CBV*, *corr. CBV*, and DS-DGE MRI maps (*LW* map from third dynamic, *AUC* maps in both grayscale and color-coded). A color-coded *AUC* calculated from the infusion block only is also shown. A graph of linewidth change versus time obtained from ROIs placed in the DS-DGE contrast-enhanced area and contralateral frontal WM is shown to the right (purple dots and orange diamonds, respectively). Regions of interest (ROIs) are overlaid on the DS-DGE MRI *AUC* map in the graph as purple and orange areas, respectively. To visualize the trend in glucose uptake, Δ*LW(t)* curves were temporally smoothed with a three-point moving average (purple and orange lines). Δ*LW*, glucose-infusion-induced *LW* change; *AUC*, area-under-curve; *corr. CBV*, corrected cerebral blood volume; *DS-DGE*, direct water saturation-dynamic glucose enhanced; *FLAIR*, fluid-attenuated inversion recovery; *Gd*, gadolinium; $K^{trans}$, volume transfer constant; *LW*, linewidth; *uncorr. CBV*, uncorrected cerebral blood volume; T$_1$w, T$_1$ weighted; $V_e$, interstitial volume.

In patients, the effect sizes for the DS-DGE *LW* differences, Δ*LW(t)*, were generally on the same order of magnitude as those obtained from simulations. This can be observed in Figure 6, where the WM uptake curve shows a small increase during and after D-Glc infusion, and GM and WM intensities in the *AUC* maps are close to zero. The same applies to the DS-DGE *AUC* maps in Figures 2, 4, and 5. CSF had a smaller Δ*LW* in the simulations than sometimes observed in the experimental data (∼15% compared to up to 40%, Figure 2). This difference may be due to the deviation from a Lorentzian line shape because of so-called sidebands appearing as distinct patterns when using short RF pulses at high sampling rates.[60] Sidebands are prone to occur in tissues with relatively long $T_2$ such as CSF or other liquid environments as necrotic tumor tissue.[60] For our simulations, the offset frequencies were selected to minimize sideband occurrence. In the experiments, if the same sidebands occur before and after infusion start, the *LW* change, however, may not be significantly affected. Interestingly, the maximum *LW* increases in blood and CSF (Figure 2) in vivo were found to be larger than those from the simulations (Figure 1). One potential explanation for blood could be an osmotic increase in blood water content due to the higher D-Glc concentration and a concomitant increase in $T_2$, previously suggested as a potential contributor to signal changes in CESL.[26,61] Another potential explanation is that the increase in plasma blood glucose concentration, and subsequently the CSF, may be greater than what we assumed in our simulations. Previous DGE studies have reported increases up to 15 mM in plasma blood glucose[13,30] for a maximum D-Glc dose of 25 g, whereas our study used 35 g. If this increase proves to be reproducible in blood, it could allow the retrieval of an arterial input function for deconvolving the tumor tissue curve, resulting in a dynamic time curve that could provide information about glucose transport and metabolism, while also removing subject variability due to insulin response. However, this is beyond the goal of this technical feasibility study.

The experimental data showed that several tumor tissues had a larger *LW* increase than GM and WM (Figures 2–6), confirming the simulations. In all four tumor cases, BBB leakage in the tumor was visible in *K2* images or when comparing uncorrected *CBV* with corrected *CBV* in areas coinciding spatially with part of the DS-DGE enhancement. However, the DS-DGE enhancement area was generally larger and more pronounced in intensity, possibly reflecting the increased sensitivity to BBB disruption of the small glucose molecule and detectability at the lower pH in tumor EES. $K^{trans}$ and $V_e$, to some degree, also showed overlap with DS-DGE enhancements, with all three parameters showing largest enhancement in the metastases. The enhancement in tumors could differ temporally, which may also reflect different amounts of BBB disruption (e.g., Figure 6). A rapid uptake would result in similar enhancement for both time periods (without or with post-infusion), while a slow uptake may show a difference.

Additional patient studies are needed to draw more substantial conclusions, and we intend to explore this in greater detail in a future study. However, these first results are highly encouraging, showing the feasibility for DS-DGE to outline tumor tissue with BBB disruption even before Gd-enhancement can detect it. Another important aspect is that the large effect in malignant tumor, combined with a small effect in normal tissue, constitutes an advantage over PET studies of brain malignancies, where phosphorylated $^{19}$F-deoxyglucose signal builds up in both brain tumors and healthy GM.[62]

There are several practical considerations that need to be mentioned. First, similar to other DGE studies, our scan time was long, which increases the risk of motion, which may introduce tissue mixing, resulting in hypointensities and hyperintensities in the DS-DGE maps.[36,37] In addition, motion can shift the voxel to another position, resulting in erroneous $\Delta LW(t)$ curves. As motion correction can reduce these errors,[35] we applied this to our dynamic data. However, care must be taken since interpolation errors can remove or reduce true effects. Furthermore, while motion correction was applied to the dynamic *LW* images, the individual frequency offset images were not corrected for motion, although these can also be affected by motion. Such a frequency-specific motion correction can be more challenging because of the intensity differences between the offsets in the dynamic Z-spectrum, especially close to the direct water saturation.[63] Similarly, partial volume effects due to motion can also induce intensity changes at individual frequencies. On the other hand, the Lorentzian fitting over a large number of points is performed using a pre-determined line shape, which may reduce small motion effects on the measured *LW*. The risk of motion can be reduced by shortening the experimental duration. However, this can introduce drawbacks such as a less robust baseline measurement and/or an incomplete glucose uptake measurement. Partial volume effects due to physiological effects, such as ventricular swelling, can also introduce signal changes.[33] Assuming CSF mixes with WM at the tissue boundary, this would result in an *LW* reduction in WM and an *LW* increase in CSF. Second, this study used a D-Glc dose of 35 g (0.5 g/kg, maximum of 35 g), while previous studies have used a maximum dose of 25 g, leading to smaller effects. For example, the first DGE MRI study at 3 T by Xu et al.[32] gave an effect size of approximately 1.5% in glioblastoma. In a larger patient cohort, Mo et al.[64] found effect sizes of 0.5% to 1.5% in low (LGG) and high grade glioma (HGG) enhancement areas using a dose of 25 g. A smaller maximum dose (20 g of D-Glc in 100 mL) was used by Bender et al.[29] in LGG and HGG patients resulting in effect sizes of maximum 0.25% in the contrast enhanced tumor area. For metastases, a recent DGE MRI study by Wu et al.[35] found up to a 10% increase after D-Glc infusion, whereas in our study, the metastases showed an increase of more than 20%. However, it is difficult to make a quantitative comparison between these studies since they differ in saturation parameters, D-Glc concentration, and contrast origin, that is, CEST including $T_2$ relaxation versus *LW* change based on $T_2$ relaxation. In addition, these differences can be caused by individual variations of the tumor structure and properties, warranting a larger population study. Third, since we are acquiring the Z-spectra during a steady state, an additional signal decrease will occur as a function of time during the Z-spectrum acquisition. Although the frequency offsets are acquired in alternating fashion around the water frequency, this continuous signal decrease will still cause the DS to deviate from a true Lorentzian line shape. This may especially affect the narrower line shapes where fitting deviations may occur at the earlier smaller intensity drops in the Z-spectrum (Figures 1 and 3). Fourth, conflicting results have been reported regarding hyperglycemia's effect on perfusion, with studies showing minimal regional changes[65] or no effect.[66,67] For our DCE and DSC measurements, we assumed the latter.

## 5 | CONCLUSION

We developed DS-DGE MRI to assess D-Glc uptake in brain tumors. Contrary to glucoCESL- and single-frequency glucoCEST-based DGE MRI, this approach is inherently independent of $B_0$ shifts occurring between the dynamics. Whole-brain dynamic Z-spectral images were acquired in less than 40 s, allowing imaging of D-Glc uptake curves in multiple tissues. Early patient data look highly promising, with DS-DGE highlighting lesion areas with information similar and/or complementary to Gd-based perfusion-weighted imaging. DS-DGE MRI can therefore further bridge the gap between research and clinical implementation of using D-Glc as a biodegradable contrast agent.

### ACKNOWLEDGMENTS
We are grateful to Terri Brawner, Kathleen Kahl, Ivana Kusevic, and Maia Lee for their assistance with the experiments. This project was supported by National Institutes of Health (RO1 EB034978 and S10OD021648), the Swedish Research Council (2019-03637), and the Swedish Cancer Society (21 1652 Pj).

### CONFLICT OF INTEREST STATEMENT
Under a license agreement between Philips and the Johns Hopkins University, L.K.'s spouse, P.C.M.v.Z., and the University are entitled to fees related to an imaging device used in the study discussed in this publication. P.C.M.v.Z. is also a paid lecturer for Philips. This arrangement has been reviewed and approved by the Johns Hopkins University in accordance with its conflict of interest policies.

### DATA AVAILABILITY STATEMENT
The data that support the findings of this study are available from the corresponding author on reasonable request.

### ORCID
*Linda Knutsson*  https://orcid.org/0000-0002-4263-113X
*Sajad Mohammed Ali*  https://orcid.org/0000-0002-4707-8206
*Anina Seidemo*  https://orcid.org/0000-0002-0919-9680

## SUPPORTING INFORMATION

Additional supporting information may be found in the online version of the article at the publisher's website.

**TABLE S1.** Exclusion criteria for the study.